\documentclass[conference]{IEEEtran}
\IEEEoverridecommandlockouts
\usepackage{cite}
\usepackage{amsmath,amssymb,amsfonts}
\usepackage{dsfont}
\usepackage{algorithmic}
\usepackage{graphicx}
\usepackage{textcomp}
\usepackage{xcolor}
\usepackage{soul}
\usepackage{algorithm}
\usepackage{algorithmic}
\usepackage{bm}
\usepackage{amsthm}
\usepackage{subfig}

\newtheorem{remark}{Remark}
\newtheorem{lemma}{Lemma}

\DeclareMathOperator*{\argmin}{\mathrm{arg\,min}}

\def\BibTeX{{\rm B\kern-.05em{\sc i\kern-.025em b}\kern-.08em
		T\kern-.1667em\lower.7ex\hbox{E}\kern-.125emX}}

\begin{document}
	\title{Integrated Communication and Control Systems:\\A Data Significance Perspective 
		\thanks{The work of S. Roth and A. Sezgin was funded by the Ministry of Economic Affairs, Industry, Climate Action and Energy of the State of North Rhine-Westphalia, Germany under grant 005-2108-0021 (5G-Expo). The work of Y. Karacora  was funded by the Ministry of Economic Affairs, Industry, Climate Action and Energy of the State of North Rhine-Westphalia, Germany under grant 005-2108-0028 (5G-Furios). The work of C. Chaccour and W. Saad was funded by the Office of Naval Research (ONR) under MURI Grant N00014-19-1-2621.}
	}
	
	\author{\IEEEauthorblockN{Stefan Roth\IEEEauthorrefmark{1}, Yasemin Karacora\IEEEauthorrefmark{1}, Christina Chaccour\IEEEauthorrefmark{3}, Aydin Sezgin\IEEEauthorrefmark{1} and Walid Saad\IEEEauthorrefmark{3}}
		\IEEEauthorblockA{\IEEEauthorrefmark{1}Ruhr University Bochum, Bochum, Germany\\
			\IEEEauthorrefmark{3}Bradley Department of Electrical and Computer Engineering, Virginia Tech, Arlington, VA, USA\\
			Email: \{stefan.roth-k21,yasemin.karacora,aydin.sezgin\}@rub.de, \{christinac,walids\}@vt.edu}}
	
	\maketitle
	
	\begin{abstract}
		The interconnected smart devices and industrial internet of things devices require low-latency communication to fulfill control objectives despite limited resources. In essence, such devices have a time-critical nature but also require a highly accurate data input based on its \emph{significance}. In this paper, we investigate various coordinated and distributed semantic scheduling schemes with a data significance perspective. In particular, novel algorithms are proposed to analyze the benefit of such schemes for the significance in terms of estimation accuracy. Then, we derive the bounds of the achievable estimation accuracy. Our numerical results showcase the superiority of semantic scheduling policies that adopt an integrated control and communication strategy. In essence, such policies can reduce the weighted sum of mean squared errors compared to traditional policies.
	\end{abstract}
	\begin{IEEEkeywords}
		Data Significance, Scheduling, ALOHA, Control, MSE, Process Monitoring
	\end{IEEEkeywords}
	
	\section{Introduction}
	
	Industrial internet of things (IIoT) devices are a staple of the fifth generation (5G) of wireless networks and beyond. The goal of IoT devices is to monitor, track, or control a system. Thereby, sensors record measurements from physical processes, and the measurements must be processed in real-time by the receiving devices. For instance, in control systems, actuators need to react immediately on variations within the measurement data. As a result, such actions must be performed in a time-critical manner while delivering accurate and error-free information \cite{9919752}. Thereby, the dynamics of the different process systems can be different. Hence, \emph{not all data should be treated equally} as the data transmitted by sensor $A$ can have a different semantics (meaning) than the ones sent by sensor $B$. As a result, such semantics can determine the state of the overall system. As a result, it is necessary to consider the levels of significance \cite{9919752,https://doi.org/10.48550/arxiv.2211.14343} and criticality requirements \cite{https://doi.org/10.48550/arxiv.2204.11878} while optimizing the overall control and communication system. This means that, latency and information freshness are not sufficient to satisfy the quality-of-service (QoS) of devices in such applications. For this reason, communication schemes should be designed to enhance information accuracy \cite{9149010} by taking the dynamics of the processes into account \cite{9919752,9691928}. Nonetheless, the communication resources of the wireless medium are limited, and thus, various resource management and allocation schemes have been proposed to orchestrate and multiplex such resources in time, space and frequency. Here, one can adopt scheduling schemes, which assign different resource blocks to demanding IoT devices. In particular, coordinated scheduling schemes such as those in \cite{9358178} and \cite{9646490} employ a common coordination among different transmitters. Note that the implementation of a central coordinator may lead to complications in practice for IIoT. Alternatively, uncoordinated or random access mechanism such as ALOHA (e.g. \cite{9162973,9488702,9132713,9605205}) do not require a central coordination. In such schemes, the decision making mechanism for transmitters is performed locally and in a distributed manner. Notably, in ALOHA schemes, packet collisions due to simultaneous access are possible, which lead to additional decoding errors compared to coordinated scheduling schemes.
	
	Thus far, in 5G systems, time critical communications have been studied in the context of so-called ultra reliable low latency communication (URLLC) paradigm. In such services, the end-to-end latency and reliability are of prime importance to be tamed under very stringent requirements. Nonetheless, recent works such as \cite{6195689} investigated the need to characterize the information freshness in contrast to the latency only. In essence, the age of information (AoI) can be evaluated to measure the information freshness of the received data at different IIoT devices. In fact, this metric has been often used to investigate the behavior of different coordinated and uncoordinated scheduling schemes \cite{9358219,9646490}. To reflect different levels of data significance, the authors of \cite{5691256} suggested the usage of \emph{semantic scheduling} to communicate over graph-structured networks. However, \cite{5691256} did not consider the presence of  the different system dynamics of the different process systems in IIoT applications. Therefore, the work in \cite{9646490} suggested to schedule each transmitter differently to enhance the estimation accuracy in terms of average mean squared error (MSE) at the receiver. However, only limited discussions are provided on how to design semantic scheduling policies for the IIoT. Clearly, investigating such schemes with an integrated control and communication lens is necessary to fully satisfy the QoS of applications with respect to data accuracy, timeliness, and freshness.
	
	\paragraph*{Contribution} 
	In this paper, we investigate different semantic scheduling strategies, which take different levels of data significance for the control performance into account. Therefore, we propose two algorithms for MSE analysis of coordinated and random access (ALOHA) scheduling schemes. Afterwards, we derive bounds of the MSE. Finally, we analyze the MSE numerically and showcase the necessity of integrating the control into scheduling schemes with the timeliness and freshness of communication via semantic schemes. Therefore, we compare the obtained regions across different strategies and show that semantic policies are beneficial.
	
	\paragraph*{Notation} The notation is as follows: Vectors, matrices and sets are denoted as bold lower-case, bold upper-case and calligraphic letters $\bm{a}$, $\bm{A}$ and $\mathcal{A}$, respectively. Superscript $\bm{A}^H$ indicates the Hermitian matrix of $\bm{A}$. $\|\bm{a}\|$ denotes the $L_2$-norm of the vector $\bm{a}$; $\mathds{E}_{a}[A(a)]$ is the expectation of $A$ over the random variable $a$. $\bm{A}\circ\bm{B}$ indicates the Hadamard product.
	
	\begin{figure}
		\centering
		\includegraphics{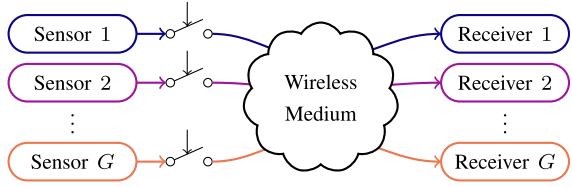}
		\caption{Multiple sensors transmit data via the wireless medium. To avoid collisions, scheduling schemes are employed.}
		\label{fig:motivation}
	\end{figure}
	
	\section{System Model and Problem Formulation}
	We consider an IoT environment, in which $G$ sensors are monitoring process systems and regularly transmitting measurements to receivers (see \figurename~\ref{fig:motivation}). As the measurements from the process systems typically change slowly over time, their values are usually modelled as linear differential matrix equation in control theory \cite{10.5555/1816978}. This means that the measurements of the $g$-th sensor $\bm{x}_g(t)$ can be described as
	\begin{align}
		\frac{\partial\bm{x}_g(t)}{\partial t}&=\bm{A}_g\bm{x}_g(t)+\bm{d}_g(t),\\
		\bm{x}_g(0)&=\bm{x}_{g,0}.
	\end{align}
	Thereby, $\bm{A}_g$ is the state-transition matrix of sensor $g$ and $\bm{d}_g(t)$ is the corresponding Gaussian-distributed input noise at time $t$, i.e., $\bm{d}_g(t)\sim\mathcal{N}(\bm{0},\bm{D}_g)$, while $\bm{x}_{g,0}$ is the measurement value at time $t=0$. Hence, the system-state value at time $t$ follows a Gauss-Markov model and is given by
	\begin{align}
		\bm{x}_g(t)&=e^{\bm{A}_gt}\bm{x}_{g,0}+\int_{0}^te^{\bm{A}_g(t-t')}\bm{d}_g(t')dt'.\label{eq:x_dynamics}
	\end{align}
	
	Each sensor $g$ measures the corresponding process values regularly at timestamps $t_{g,i}\forall i$ according to a channel access strategy $S_g=(t_{g,1},t_{g,2},\dots)$ and transmits the measurements directly afterwards. The different levels of significance are considered within the channel access strategies, as we will elaborate on later. In coordinated scheduling, the $G$ different channel access strategies are correlated (due to central coordination), while they are uncorrelated in random access schemes such as ALOHA. Each transmission requires a time $\Delta$ to be finished. When two sensors schedule their packets to be transmitted partially simultaneously, the packets collide. In this case, none of the packets can be decoded successfully. Otherwise, there is a probability $\varepsilon$, with which channel noise will lead to an unsuccessful decoding. When a data packet with index $i$ is decoded successfully, its data are used to estimate the system-state value $\bm{x}_g(t)$ until a following packet is decoded successfully. From \eqref{eq:x_dynamics}, the estimate can be obtained as
	\begin{align}
		\bm{\hat{x}}_g(t)&=e^{\bm{A}_g(t-t_{g,i})}\bm{x}_g(t_{g,i}). \label{eq:x_estimate}
	\end{align}
	The time-span in which each data packet is used at the receiver for estimation can be described by an AoI interval $[\underline{\tau},\overline{\tau})$. Thereby, the AoI bounds are $\underline{\tau}=\Delta$ and $\overline{\tau}$ and might vary for each packet and depend on the channel access strategy.
	
	The accuracy of the information at the receiver can be described by the MSE between the estimate of the $g$-th receiver and the corresponding system-state, i.e.,
	\begin{align}
		\mathsf{MSE}_g=\mathds{E}_{t}\left[\left\|\bm{\hat{x}}_g(t)-\bm{x}_g(t)\right\|^2\right].\label{eq:mse}
	\end{align}
	Note that the measurements from the process systems are ergodic. From \cite{7524524}, \eqref{eq:mse} can be casted as fraction of the average packet-integrated MSE and the average time-span over which a certain packet is used for packet estimation, i.e.,
	\begin{align}
		\mathsf{MSE}_g &= \frac{\mathds{E}_{\underline{\tau},\overline{\tau}}\left[L_g\left(\underline{\tau},\overline{\tau}\right)\right]}{\mathds{E}_{\underline{\tau},\overline{\tau}}\left[\overline{\tau}-\underline{\tau}\right]}.\label{eq:MSEwithL}
	\end{align}
	Here, $L_g\left(\underline{\tau},\overline{\tau}\right)$ is the packet-integrated MSE, i.e., the integral of the instantaneous MSE over the AoI interval $[\underline{\tau},\overline{\tau})$. This value is described by Lemma~\ref{lem:packet-integrated-MSE} that follows directly from \cite{9646490}.
	\begin{lemma}\label{lem:packet-integrated-MSE}
		The packet-integrated MSE equals
		\begin{align}
			L_g\left(\underline{\tau},\overline{\tau}\right)&=\mathrm{trace}\Big\{e^{\bm{A}_g\overline{\tau}}\bm{\Phi}_ge^{\bm{A}_g^H\overline{\tau}}-e^{\bm{A}_g\underline{\tau}}\bm{\Phi}_ge^{\bm{A}_g^H\underline{\tau}}\nonumber\\&\hspace{3.9cm}-\bm{\Upsilon}_g\left(\overline{\tau}-\underline{\tau}\right)\Big\},\label{eq:packet-integrated-mse}
		\end{align}
		where
		\begin{align}
			\bm{\Upsilon}_g&=\bm{U}_g\left(\left(\bm{U}_g^{-1}\bm{D}_g\bm{U}_g^{-H}\right)\circ\bm{B}\right)\bm{U}_g^H,\\
			\bm{\Phi}_g&=\bm{U}_g\left(\left(\bm{U}_g^{-1}\bm{\Upsilon}_g\bm{U}_g^{-H}\right)\circ\bm{B}\right)\bm{U}_g^H.
		\end{align}
		Moreover, $\bm{A}_g=\bm{U}_g\bm{\Lambda}_g\bm{U}_g^{-1}$ is the eigenvalue decomposition of $\bm{A}_g$ and $\left(\bm{B}\right)_{m,n}=\big(\left(\bm{\Lambda}_{g}\right)_{m,m}+\left(\bm{\Lambda}_{g}\right)_{n,n}^*\big)^{-1}$.
	\end{lemma}
	\begin{remark}
		In \eqref{eq:packet-integrated-mse}, the exponents of the first term, i.e., $\bm{A}_g\overline{\tau}$, describe the product of the state-transition matrix $\bm{A}_g$ and the peak AoI $\overline{\tau}$. This shows that the MSE of an estimate with index $g$ is low, if either the AoI is kept small or if the corresponding $\bm{A}_g$ has small eigenvalues. As the eigenvalues of the $G$ state-transition matrices can have very different amplitudes, a frequent transmission is not equally important for all sensors. In other words, the data transmitted by the different sensors have different levels of significance.
	\end{remark}
	
	We now recall that the channel access strategies $S_g$ of the different sensors $g\in\{1,\dots,G\}$ can be correlated (in the case of coordinated scheduling setups) or uncorrelated (in the case of ALOHA setups). For both cases, our target is the selection of channel access strategies from a set of pre-defined channel access strategies $\mathcal{S}$, such that the weighted MSE is minimized, i.e.,
	\begin{align}
		\underset{S_g\in\mathcal{S}, g=1,\dots,G}{\mathrm{\ minimize\ }} \sum_{g=1}^G\alpha_g \mathsf{MSE}_g.\label{eq:optProblem}
	\end{align}
	In \eqref{eq:optProblem}, $\alpha_g, g=1,\dots,G$ are weight factors that indicate different levels of criticality, which are defined such that $\sum_{g=1}^G\alpha_g=1$. In the following, we will discuss different channel access strategies. and then provide algorithms to evaluate a fixed set of channel access strategies. The channel access strategies selected will then be the ones which minimize the MSE.
	
	\section{Channel Access Strategies for Semantic Scheduling}
	Various channel access strategies can be applied to coordinated and random access scheduling. In this section, we will highlight how semantic scheduling strategies enable an integrated control and communication strategy that addresses data significance.
	
	\subsection{Channel access strategies in coordinated scheduling setups}
	For the case of coordinated scheduling, we consider the following channel access strategies (see also \figurename~\ref{fig:scheduling_illustration}):
	\begin{figure}
		\centering
		\subfloat[max-trials-$(\infty,1)$]{\includegraphics{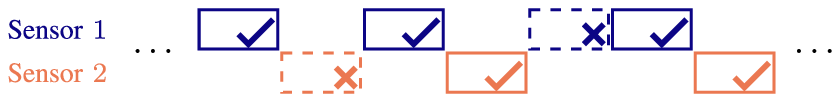}}\\
		\subfloat[multiple-success-$(2,1)$]{\includegraphics{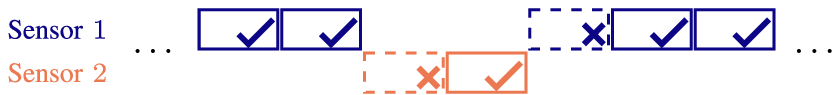}}
		\caption{Illustration of the different coordinated scheduling policies for two sensors. The transmissions of the two sensors are shown in blue and orange; ticks and crosses indicate successful and unsuccessful decoding, respectively.}
		\label{fig:scheduling_illustration}
	\end{figure}%
	\begin{figure}
		\centering
		\subfloat[individual-CAP]{\includegraphics{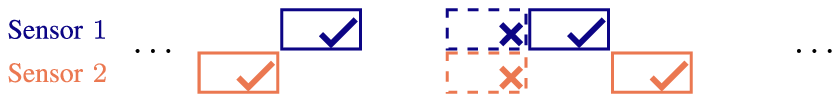}}\\
		\subfloat[threshold-ADRA]{\includegraphics{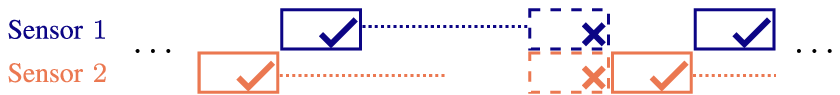}}\\
		\caption{Illustration of the different ALOHA policies for two sensors. The transmissions of the two sensors are shown in blue and orange; ticks and crosses indicate successful and unsuccessful decoding, respectively.}
		\label{fig:aloha_illustration}
	\end{figure}
	\begin{itemize}
		\item In \cite{9646490}, a semantic scheduling policy was proposed, which we refer to as \emph{max-trials-$(P_1, P_2,\dots)$}. Thereby, the sensors alternately transmit packets. Each sensor $g$ has at most $P_g$ trials, but finishes once one transmission is successful. Then, the following sensor starts transmitting. In the special cases max-trials-$(1, 1,\dots)$ and max-trials-$(\infty, \infty,\dots)$, this policy becomes equal to the widely used round-robin and maximum-age scheduling policies.
		\item We further propose a novel semantic scheme referred to as \emph{multiple-success-$(Q_1,Q_2,\dots)$}. Thereby, the total considered time is divided into intervals. Within each interval, all sensors are scheduled such that each sensor $g$ transmits within $Q_g$ transmission blocks. Within each transmission block, the sensor transmits packets until one packet is received successfully. The $Q_1+\dots+Q_G$ transmission blocks are ordered such that the transmission blocks of each sensor are separated as much as possible. Hence, the different levels of significance are relevant even if there are no decoding errors.
	\end{itemize}
	\subsection{Channel access strategies in ALOHA setups}
	In ALOHA schemes, each sensor applies an own channel access strategy, in which the transmit times are stochastically distributed. While pure-ALOHA schemes do not employ any coordination between the sensors, slotted-ALOHA schemes allow the agreement on specific slots that can be used for transmission. Within our analysis, we focus on slotted-ALOHA. The following schemes are considered (see \figurename~\ref{fig:aloha_illustration}):
	\begin{itemize}
		\item Each sensor $g$ can have an individual channel access probability (CAP) $R_g$ \cite{9162973}. Over all slots, the CAPs are constant for each sensor. However, we consider CAPs that can vary across the different sensors and thus take differnet levels of significance into account. We refer to this scheme as \emph{individual-CAP-$(R_1,R_2,\dots)$}.
		\item Moreover, a threshold-based distributed age-dependent random access (ADRA) \cite{9162973} scheme is analyzed. Thereby, each sensor is pausing after a successful reception until a certain AoI $\tilde{\tau}_g$ is reached. Afterwards, packets are transmitted in each slot with the CAP $R_g$ until a successful reception. Again, we assume that individual $\tilde{\tau}_g$ and $R_g$ can be chosen for each sensor, which take the different levels of significance into account. We refer to this scheme as \emph{threshold-ADRA-$(R_1,R_2,\dots)$-$(\tilde{\tau}_1,\tilde{\tau}_2,\dots)$}.
	\end{itemize}

	\section{MSE Analysis}
	
	For each considered set of channel access strategies $S_1,\dots,S_g$, the time-average MSE can be computed. Here, we have to distinguish between the cases of coordinated scheduling strategies and distributed channel access strategies (as in ALOHA setups).
	
	\subsection{MSE in coordinated scheduling setups}
	Our target is now to develop an algorithm that simulates a large number $K$ of packet transmissions to obtain the time-average MSE in coordinated scheduling setups. Thereby, we determine the expected MSE numerically by integrating the MSE over all packet transmissions and dividing the result by the total considered time-span. Recall that the transmit times are chosen in a centralized way, such that collision errors are avoided. Hence, decoding errors can only occur due to receiver noise, such that the error probability is $\varepsilon$ and constant for all packets. Once a data transmission is simulated to be successful, we can calculate the packet-integrated MSE and update the transmit time and transmission delay of the data at the receiver. 
	\begin{algorithm}[tb]
		\caption{Average MSE with centralized scheduling}
		\begin{algorithmic}
			\STATE \textbf{Input:} Number of packets simulated $K$, channel access strategies $S_1,\dots,S_g$
			\STATE Initialize $\underline{\tau}_g\gets 0$, $t_g\gets 0$, $\underline{t}_g\gets 0$, $\ell_g\gets 0$ for all $g$
			\FOR{$k\gets1$ \textbf{to} $K$}
			\STATE $(t_g,g)\gets$ time and source scheduled by $S_1,\dots,S_G$
			\STATE $\Delta\gets$ (possibly stochastic) transmit time
			\IF{rand() $<1-\varepsilon$}
			\STATE $\ell_g\gets \ell_g + L_g(\underline{\tau}_g,t_g+\Delta-\underline{t}_g)$\hfill\COMMENT{integrate~MSE}
			\STATE $\underline{t}_g\gets t_g$ \hfill\COMMENT{store transmit time}
			\STATE $\underline{\tau}_g\gets \Delta$ \hfill\COMMENT{store transmission delay}
			\ENDIF
			\ENDFOR
			\STATE $t_{g'}\gets t_g$ for all $g'$
			\STATE $\ell_{g'}\gets \ell_{g'} + L_{g'}(\underline{\tau}_g,t_{g'}-\underline{t}_{g'})$ for all $g'$ \hfill\COMMENT{integrate~MSE}
			\STATE \textbf{Output:} $\mathsf{MSE}_{g'}\gets \ell_{g'}/t_{g'}$ for all $g'$\hfill\COMMENT{Output MSE}
		\end{algorithmic}
		\label{alg:schedulingMseAoI}
	\end{algorithm}
	At the end, the MSE values are updated for all receivers, such that equal time intervals are considered. The full algorithm is shown in Algorithm~\ref{alg:schedulingMseAoI}, where rand() generates a continuous uniform distributed variable on $[0,1)$.
	
	With this algorithm, strategies involving a coordinated access scheme can be evaluated. To evaluate strategies with an uncoordinated access, the algorithm needs to be modified.
	
	\subsection{MSE in ALOHA setups}
	Herein, we propose an algorithm to analyze the MSE in ALOHA setups. The MSE is again evaluated via simulations over a large number $K$ of transmitted packets. Different to the above, the sensors schedule their packets independently. Thus, in addition to receiver noise, collisions of the transmissions can also lead to decoding errors. Hence, all channel access strategies need to be modeled simultaneously to detect, which sensor is transmitting first and which packets collide. Hence, we store the next transmit time of each sensor during the entire algorithm. Within each loop iteration, we then simulate the transmission of the earliest packet scheduled by any of the sensor. The packet decoding is assumed successful, if all of the following three conditions hold:
	\begin{itemize}
		\item First, there must be no other sensor scheduled to start a transmission within the transmission interval of the packet. This definition is mathematically equivalent to the condition $\nexists g'\neq g:t_{g'}<t_g+\Delta$.
		\item Second, there must be no other sensor scheduled to start a transmission shortly in advance, such that the transmissions overlap. This condition is equal to the first condition in the previous iteration. In the proposed algorithm, we use an additional variable $f$ to evaluate this in the previous iteration.
		\item Third, also the receiver noise must not lead to decoding errors. This condition is fulfilled, if rand() $<1-\varepsilon$, where rand() refers to a uniform distributed variable on $[0,1)$.
	\end{itemize}
	If all conditions hold for a packet, the integrated MSE is incremented by the packet-integrated MSE. Additionally, the transmit time and the transmission delay of the received data are updated. At the end of the iteration, the variable $f$ is updated such that collisions are recognized. Further, the scheduled sensor obtains its next transmission time from the according channel access strategy. The full algorithm is presented in Algorithm~\ref{alg:alohaMseAoI}.
	\begin{algorithm}[tb]
		\caption{Average MSE and AoI with distributed ALOHA}
		\begin{algorithmic}
			\STATE \textbf{Input:} Number of packets simulated $K$, channel access strategies $S_1,\dots,S_g$
			\STATE Initialize $\underline{\tau}_g\gets 0$, $t_g\gets 0$, $\underline{t}_g\gets 0$, $\ell_g\gets 0$ for all $g$, $f\gets 1$
			\STATE $t_g\gets$ time scheduled by $S_g$ for all $g$
			\FOR{$k\gets1$ \textbf{to} $K$}
			\STATE $g\gets\argmin_{g}(t_1,\dots,t_G)$ \hfill\COMMENT{source of next packet}
			\STATE $\Delta\gets$ (possibly stochastic) transmit time
			\IF{$\nexists g'\neq g:t_{g'}<t_g+\Delta$ \textbf{and} $f$ \textbf{and} rand() $<1-\varepsilon$}
			\STATE $\ell_g\gets \ell_g + L_g(\underline{\tau}_g,t_g+\Delta-\underline{t}_g)$\hfill\COMMENT{integrate~MSE}
			\STATE $\underline{t}_g\gets t_g$ \hfill\COMMENT{store transmit time}
			\STATE $\underline{\tau}_g\gets \Delta$ \hfill\COMMENT{store transmission delay}
			\ENDIF
			\STATE $f\gets \nexists g'\neq g:t_{g'}<t_g+\Delta$
			\STATE $t_g\gets$ time scheduled by $S_g$
			\ENDFOR
			\STATE $t_{g'}\gets t_g$ for all $g'$
			\STATE $\ell_{g'}\gets \ell_{g'} + L_{g'}(\underline{\tau}_g,t_{g'}-\underline{t}_{g'})$ for all $g'$ \hfill\COMMENT{integrate~MSE}
			\STATE \textbf{Output:} $\mathsf{MSE}_{g'}\gets \ell_{g'}/t_{g'}$ for all $g'$\hfill\COMMENT{Output MSE}
		\end{algorithmic}
		\label{alg:alohaMseAoI}
	\end{algorithm}
	\section{Upper and Lower Bounds of the MSE}\label{sec:boundary}
	In this section, we investigate the bounds of the MSE of the estimate at receiver $g$. These reference values enable the evaluation of the quality of the considered strategies. From the case where the sensor $g$ schedules transmissions sequentially without any waiting time, while other sensors are silent (such that no collisions occur), the MSE \emph{is lower bounded}.  Based on the distributions of the number of unsuccessful transmissions $N$ between two successful transmissions and the transmit times $\Delta_n$, the MSE can be obtained from \eqref{eq:MSEwithL} as
	\begin{align}
		\mathsf{MSE}_{g,\mathrm{LB}}&=\frac{\mathds{E}_{\Delta_n\forall n,N}\left[L_g\left(\Delta_1,\sum_{n=1}^{N+2}\Delta_n\right)\right]}{\mathds{E}_{\Delta_n\forall n,N}\left[\sum_{n=2}^{N+2}\Delta_n\right]}.
	\end{align}
	If the distribution of the transmit time is known, this value can be calculated analytically or numerically. If $\Delta_n$ is time-constant, i.e., $\Delta_n=\Delta$, this lower-bound becomes
	\begin{align}
		\mathsf{MSE}_{g,\mathrm{LB}}&\nonumber\\&\hspace{-0.8cm}=\frac{\mathrm{trace}\left\{e^{\bm{A}_g2\Delta}\bm{\Psi}_ge^{\bm{A}_g^H2\Delta}-e^{\bm{A}_g\Delta}\bm{\Phi}e^{\bm{A}_g^H\Delta}-\frac{\Delta}{1-\varepsilon}\bm{\Upsilon}_g\right\}}{\frac{\Delta}{1-\varepsilon}},
	\end{align}
	in which
	\begin{align}
		\bm{\Psi}_g&=\bm{U}_g\left(\left(\bm{U}_g^{-1}\bm{\Phi}_g\bm{U}_g^{-H}\right)\circ\bm{C}_g\right)\bm{U}^H,\\
		\left(\bm{C}_g\right)_{m,n}&=\frac{1-\varepsilon}{1-\varepsilon e^{\left(\lambda_{g,m}+\lambda_{g,n}^*\right)\Delta}}.
	\end{align}
	
	An \emph{upper bound} can be obtained for the case that all transmitted packets collide. If the system is stable (i.e., if all eigenvalues of $\bm{A}_g$ are negative), the upper-bound will be finite and can be obtained from \eqref{eq:MSEwithL} for the case of $\overline{\tau}\rightarrow\infty$. For unstable systems instead (i.e., where $\bm{A}_g$ has any positive eigenvalues), the upper-bound will be infinite.
	\begin{align}
		M_{g,\mathrm{UB}}&=\begin{cases}-\mathrm{trace}\left\{\bm{\Upsilon}\right\}&\mathrm{eig}_i\left(\bm{A}_g\right)<0\forall i\\\infty&\text{otherwise}\end{cases}.
	\end{align}
	In the following we will analyze the effects of the channel access strategies on the MSE numerically and compare their MSE values to the bounds.

	\begin{figure}
		\centering
		\includegraphics{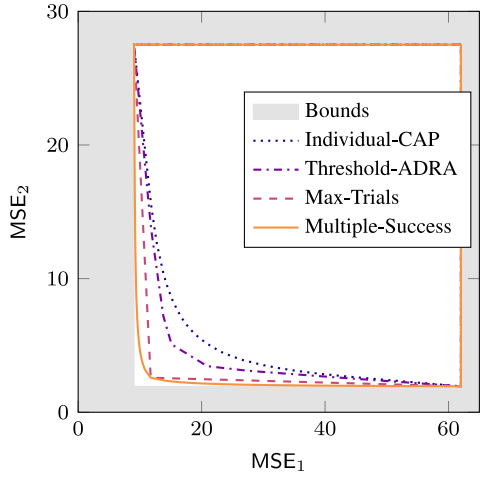}
		\caption{Achievability regions of the MSE values obtained with the different scheduling strategies for the stable systems.}
		\label{fig:scheduling_stable}
	\end{figure}
	
	\section{Numerical Results}
	We consider IoT environments with $G=2$ sensors and investigate the MSE of the estimates at both receivers. Thereby, we first focus on systems with stable dynamics, i.e., negative eigenvalues, before then considering systems with unstable dynamics, i.e., positive eigenvalues. In both cases, the channel is characterized by $\Delta=1$ and $\varepsilon=0.05$.
	\subsection{Achievability region for stable systems}
\begin{figure}
		\centering
		\includegraphics{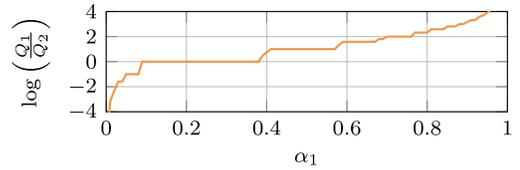}
		\caption{Parameters of multiple-success for the stable systems.}
		\label{fig:scheduling_stable_parameters}
	\end{figure}

	We first consider two stable systems\footnote{We consider two of the process systems from \cite{9646490}, where the first has parameters
		\begin{align}
			\bm{A}_1&=\begin{pmatrix}-0.04&0.03&-0.05\\-0.01&-0.06&0.05\\0.2&0.15&-0.4\end{pmatrix},&
			\bm{D}_1&=\begin{pmatrix}4&1&3\\1&0.25&0.75\\3&0.75&2.25\end{pmatrix}\nonumber
		\end{align}
		and the second system is specified by
		\begin{align}
			\bm{A}_2&=\begin{pmatrix}-0.02&0\\0&-0.03\end{pmatrix},&\bm{D}_2&=\begin{pmatrix}0.7&0.2\\0.2&0.6\end{pmatrix}.\nonumber
	\end{align}}, i.e., $\bm{A}_1$ and $\bm{A}_2$ have solely negative eigenvalues, such that the MSE will stay bounded (even if no data are transmitted). In \figurename~\ref{fig:scheduling_stable}, we consider the achievability regions of the MSEs obtained with the different scheduling policies. The parts of the lines on the bottom and the left are Pareto-optimal for the according class of policies. To obtain the regions for all policies, we have assumed that time-sharing between the obtained points is possible. As expected, the results show that the policies involving central scheduling outperform the random access policies. Note that the MSE regions, which can be obtained by the traditional scheduling policies round-robin and maximum-age are almost identical to the regions of the max-trials policy (and thus are not shown). However, the multiple-success policy is able to fill significant parts of the area between the aforementioned policies and the boundary from section~\ref{sec:boundary}. \figurename~\ref{fig:scheduling_stable_parameters} shows that, with this policy, for most weights $\alpha_1$, a significance-based choice of parameters for the two sensors is optimal. 
	\begin{figure}
		\centering
		\includegraphics{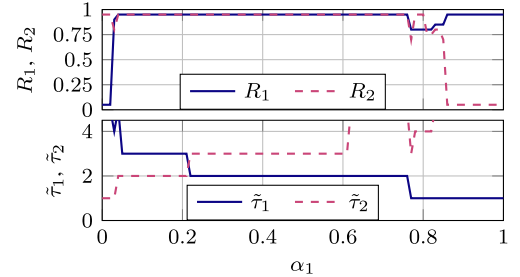}
		\caption{Parameters of threshold-ADRA for the stable systems.}
		\label{fig:aloha_stable_parameters}
	\end{figure}
	\figurename~\ref{fig:scheduling_stable} also shows that the threshold-ADRA policy outperforms the individual-CAP policy. The optimal parameter choice of threshold-ADRA is shown in \figurename~\ref{fig:aloha_stable_parameters}. The plot shows that also here a significance-based parameter choice is beneficial, especially significance-based thresholds.

	\subsection{Achievability region for unstable systems}
	
\begin{figure}
		\centering
		\includegraphics{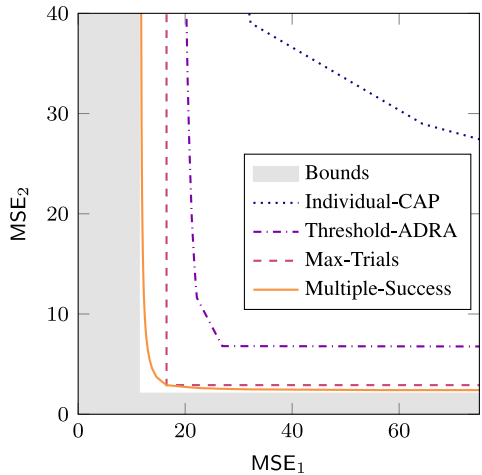}
		\caption{Achievability regions of the MSE values obtained with the different scheduling and ALOHA strategies for the unstable systems.}
		\label{fig:scheduling_unstable}
	\end{figure}

\begin{figure}
		\centering
		\includegraphics{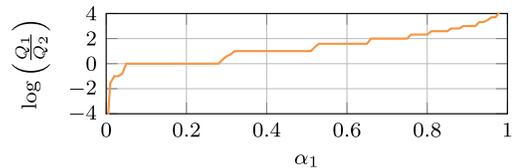}
		\caption{Parameters of multiple-success for the unstable systems.}
		\label{fig:scheduling_unstable_parameters}
	\end{figure}

\begin{figure}
		\centering
		\includegraphics{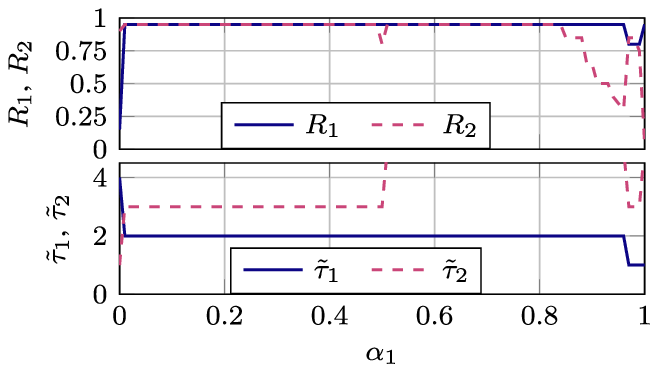}
		\caption{Parameters of threshold-ADRA for the unstable systems.}
		\label{fig:aloha_unstable_parameters}
	\end{figure}

	To investigate systems with unstable dynamics, we now employ slightly different parameters\footnote{We now modify the system parameters of the previous section such that $\bm{A}_1'=-\bm{A}_1$ and $\bm{A}_2'=-\bm{A}_2$. The matrices $\bm{D}_1'$ and $\bm{D}_2'$ remain as before, i.e., $\bm{D}_1'=\bm{D}_1$ and $\bm{D}_2'=\bm{D}_2$.}, within which the system matrices have positive eigenvalues. The regions of the achievable MSE values with the considered scheduling and ALOHA policies are shown in \figurename~\ref{fig:scheduling_unstable}. Compared to the previous results, the MSE is not bounded and the gaps between the policies and the lower-bound are larger. The coordinated scheduling policies also here outperform the ALOHA schemes. Yet, also in the case of the considered unstable systems, the traditional policies round-robin and maximum-age again have similar achievability regions as the max-trials policy. Similarly to stable systems, the multiple-success policy is able to fill the gap among the aforementioned policies and the bounds of the MSE significantly. \figurename~\ref{fig:scheduling_unstable_parameters} shows that here, even for a wider range of $\alpha_1$, a significance-based parameter choice is beneficial. The achievability region of the different ALOHA policies with this system is also shown in \figurename~\ref{fig:scheduling_unstable}. Thereby, all parts of the region obtained with the individual-CAP probability contain only relatively high MSE values. Compared to this, the threshold-ADRA policy leads to a significantly enlarged achievability region. For this policy, almost all Pareto-optimal points can only be achieved due to a significance-based parameter choice (see \figurename~\ref{fig:aloha_unstable_parameters}).
	
	\section{Conclusion}
	Traditionally, control and communication systems have been optimized separately. In this paper, we have presented two algorithms to investigate the impact of the medium access design, i.e., coordinated scheduling and ALOHA, on the usability for control systems in terms of MSE. The numerical results show that semantic scheduling, i.e., assigning a higher priority to some of the sensors than to others due to different levels of data significance may be superior in stable and unstable dynamic systems. By doing so, the MSE of the estimate at one receiver can be enhanced on the cost of reducing the MSE at another receiver. As priorities and system dynamics can be very different, the overall usability of the data for certain applications can be increased. 
	
	
	\bibliographystyle{IEEEtran}
	\bibliography{main}
	
\end{document}